# Semantic Annotation: The Mainstay of Semantic Web

Thabet Slimani
Taif University
Taif, Saudia Arabia

Thabet.slimani@gmail.com

**Abstract**:

Given that semantic Web realization is based on the critical mass of metadata accessibility and the representation of data with formal knowledge, it needs to generate metadata that is specific, easy to understand and well-defined. However, semantic annotation of the web documents is the successful way to make the Semantic Web vision a reality. This paper introduces the Semantic Web and its vision (stack layers) with regard to some concept definitions that helps the understanding of semantic annotation. Additionally, this paper introduces the semantic annotation categories, tools, domains and models.

## 1. INTRODUCTION

Currently, despite the large amounts of documents and resources available online, semantic analysis is not enough supported by internet search engines (they typically match words syntactically). However, the requirement of massive metadata for the web content allows various Semantic Web applications to appear and gain broad approval. A typical Web application would provide and use new access methods based on the associated metadata. The Semantic Web was devised by Tim Berners-Lee as a network that includes a content semantically-enriched which contains links to explicit, formal semantics. A good number of the semantic content available has been generated automatically by the mean of wrapping or by using annotation services. However, Semantic Web success depends on the accomplishment of a great number of users creating and exploiting semantic content. This achievement requires tools that reduce the complexity of semantic technologies.

Various IE technologies are currently available allowing named entity recognition within the text, events, relations and scenarios in which they exist. Metadata assigned to a document can range from author reference of the document, to annotations of all the entities referred to in the text. To make this metadata readable by machines for effective structuring, discovery, automation, integration, and reuse is an important issue in semantic research. Based on the category of annotation, the automatic (versus manual) extraction of metadata approach is scalable, author-independent, and not expensive and enriches the web content of a specific user. At present, the technology available to provide automatic semantic annotation is not yet mature to achieve intuitive, scalable, and accurate model for generation and representation of such annotations.

This paper presents first a comprehensive introduction to Semantic Web (layers and content enrichment of web resources). Semantic annotation categories (manual, semi-automatic and automatic annotation) are presented in sections 3. Section 4 offers a brief explanation of the semantic annotation models and domains. Some existing annotation tools classified by text, images and ontologies are described in the section 5. Finally, section 6 gives a conclusion with perspectives for future work.

## 2. SEMANTIC WEB INTRODUCTION

The Semantic Web is a vision created by Tim Berners-Lee, the inventor of the WWW [1]. The success of the current WWW leads to a new challenge: A huge number of data is only human understandable; machine support is limited or absent. Berners-Lee suggests mechanisms to describe data in Semantic Web terms which will facilitate applications to exploit data (machine processable) in more ways and support the user in his task. The relevant pages and can thus improve both precision and recall. The definition of a Semantic Web structure is crucial. The structure has to be defined, and then has to be filled with life. To do this task, one should start with the easier tasks first. The following steps show the direction where the Semantic Web is heading:

- Provide a common syntax for machine understanding.
- Create common vocabularies.
- Support logical language.
- Use the language for exchanging proofs.

The layer structure of the Semantic Web suggested by Tim Berners-Lee reflects the previous steps that follow the understanding that each step alone will already add value, so that the Semantic Web can be implanted in an incremental approach.

### 2.1 Layers of Semantic Web

The layers of Semantic Web suggested by Berners-Lee is a stack which shows how technologies that are standardized for the Semantic Web are organized to make the Semantic Web possible. This architecture is discussed in detail in [2] and [3], which also address recent research issues (Figure 1):

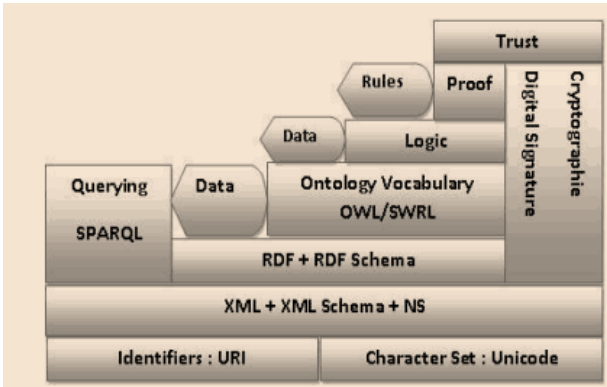

Figure. 1 The layers of Semantic Web.

The bottom layers contain technologies providing common syntax. Uniform Resource Identifier (URI) provides means for uniquely identifying semantic web resources (entities)[1], while Unicode serves to represent and manipulate text in many languages useful for exchanging symbols. The Extensible Markup Language (XML) is a markup language that enables creation of documents composed of structured data, and XML Schema allows the definition of grammars for valid XML documents. Semantic web gives meaning (semantics) to structured data. XML documents can refer to different namespaces to make explicit the context (and therefore meaning) of different tags. XML Namespaces provide a way to use markups from more sources. Semantic Web aims to connect data together, which needs to refer more sources in one document.

The explained two layers are nowadays broadly accepted, and the number of XML documents is growing quickly. XML is the first step in the right direction, but it only formalizes the structure of a document and not its content. The Resource Description Framework (RDF)[2] is a framework for creating statements in a form denoted by triples. This form enables the representation of information about resources in the form of graph and can be seen as the first layer where information becomes machine understandable: According to the W3C recommendation[3], RDF "is a foundation for processing metadata; it provides interoperability between applications that exchange machine understandable information on the Web". The components of each RDF document consist of three types of entities: Resources (subjects and objects), properties (predicates/relations). Resources represent Web pages, parts or set of Web pages, or anything (real-world object) that can have a URI. Properties are specific attributes, or relations describing resources. The combination of a resource together with a property having a value for that resource forms a Statement (known as the subject, predicate and object). A value is either a literal, a resource, or other statement. A Statement which may be represented as a triple of the form (Subject, Property, Object) asserts that a resource recognized by the subject, has a property whose value is the recognized by the object (either another resource or a literal). Consequently, a property is a binary relationship between two resources or between a resource and a literal value. Figure 2 shows an example of RDF statements. Two of the Researcher of the "ASSW project" (i.e., their Web pages) are represented as resources 'URI-Mam' and 'URI-Ram'. On the lower right of Figure 2, the statement consists of the resource 'URI-Mam' and the property 'cooperates-with' with the value 'URI-Ram' (resource). The resource 'URI ASSW' has as value for the property 'title' the literal "Annotation System for Semantic Web".

RDF is basically a directed graph with labelled edges and partially labelled nodes. The definition of a simple modelling language on top of RDF is realized by the RDF Schema (RDFS)[4] which includes classes, IS-a relationships between classes and properties, and properties characterized by domain/range restrictions. RDF and RDF Schema are structured in XML syntax, but they do not use the tree semantics of XML. An extension of RDFS including more advanced constructs to describe semantics of RDF statements based on description logic is provided by Web Ontology Language (OWL)[4]. It allows states additional constraints, such as for example cardinality, value restrictions, or characteristics of properties such as transitivity. The ontology vocabulary denotes the next layer. Gruber [5], define an ontology as "an explicit formalization of a shared understanding of a conceptualization". Most of the definitions realized by different research communities share a certain understanding in common: That means, ontology is a formal model which explicitly represents the consensual knowledge of a domain. The domain entities are modelled through a set of concepts, a hierarchy on them, and relations between concepts. By instantiating these ontological concepts, concrete facts and information items which can be stored in the ontology. Most of these definitions also include axioms in some specific logic. The core of own ontology definition is presented in the following section.

At the layer of ontology vocabulary it is possible to query any RDF-based data (i.e., including statements involving RDFS and OWL) with the use of the latest RDF query language (SPARQL) [6]. According to Berners-Lee, the next layer is Logic. Nowadays, the integration between ontology and the logic levels is treated by the most researchers. This integration is encouraged by the ability of the most ontologies to allow for logical axioms. With the applicability of logical deduction, we can infer new knowledge from the information which is stated explicitly. For instance, the axiom given above allows to logically infer that the researcher addressed by 'URI-RAM' cooperates with the researcher addressed by 'URI-MAM'. The feasibility of the type of inference depends deeply on the logics chosen.

---

[1] Refers to a locatable URI, e.g., an http://www.w3schools.com/RDF address. It is often used as a synonym, although URLs are a subclass of URIs, see http://www.w3.org/Addressing
[2] http://www.w3.org/TR/REC-rdf-syntax/
[3] http://www.w3.org/TR/REC-rdf-syntax-grammar-20040210/
[4] RDF Vocabulary Description Language 1.0: RDF Schema. 2004 Available from: http://www.w3.org/TR/rdf-schema/.

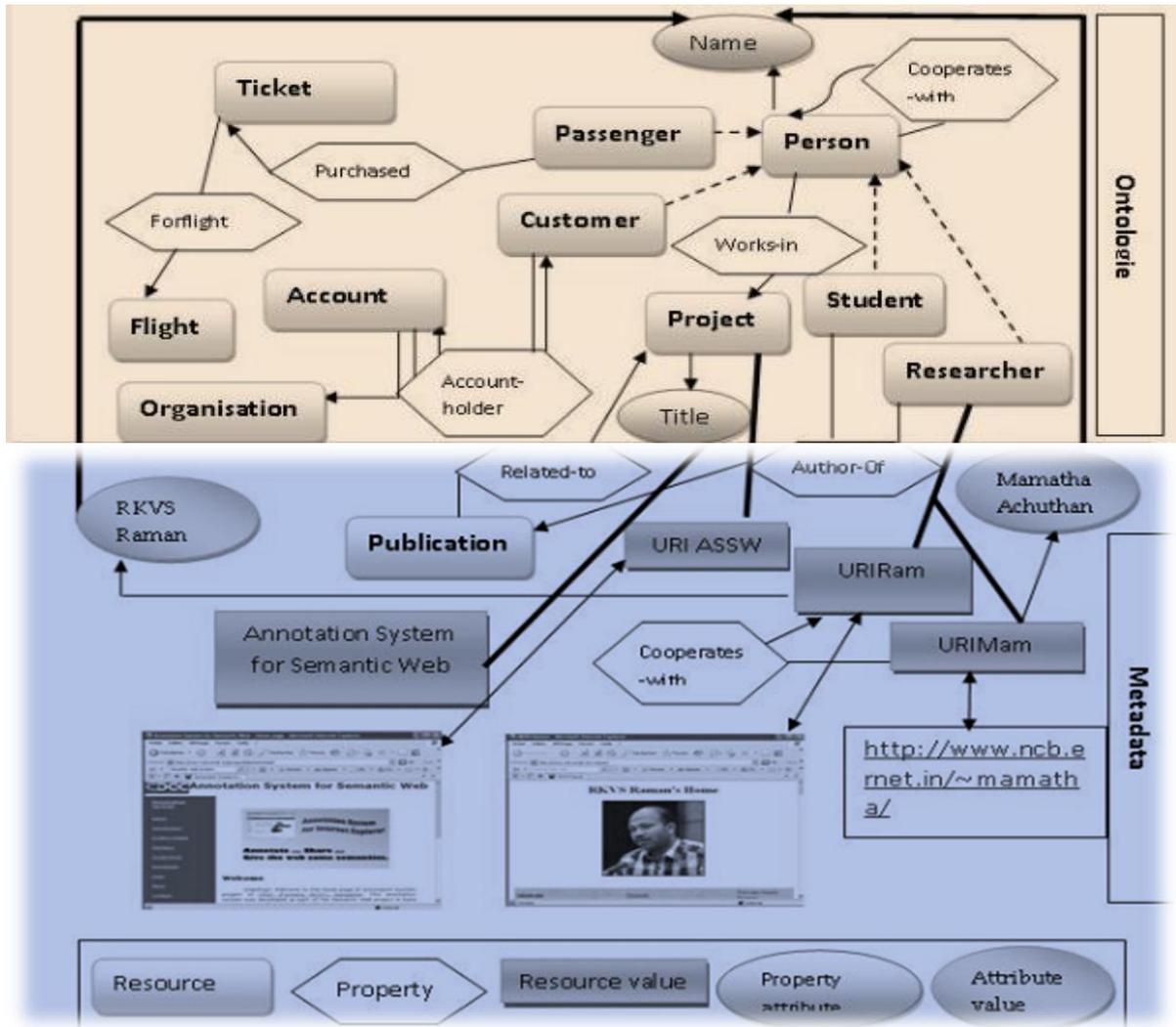

Figure. 2 The relation between resources in RDF graph and ontology.

The remaining layers are **Proof and trust**. The top layers contain technologies that are not yet standardized which require the ability to check the validity of the statements made in the (Semantic) Web, and Trust to derive statements will be supported by (a) verifying that the premises come from trusted sources and by (b) relying on formal logic during deriving new information. Consequently, the way it processes information will increase in the presence of validated statements. Then, the author must provide a proof which should be provable by a machine. At this level, it is not required that the machine of the reader finds the proof, but he needs only to check the proof given by the author. These two layers are rarely undertaken in current research.

## 3. SEMANTIC ANNOTATION CATEGORIES

Semantic annotation is the process that creates semantic labels of documents for the semantic Web, aiming to support advanced searching (based on concepts), reasoning about Web resources and the information visualization based on ontology. Additionally annotation is used to convert syntactic structures into knowledge structures. In other terms, semantic annotation consists to generate specific metadata and usage schema, enabling new information access methods and extending the existing ones.

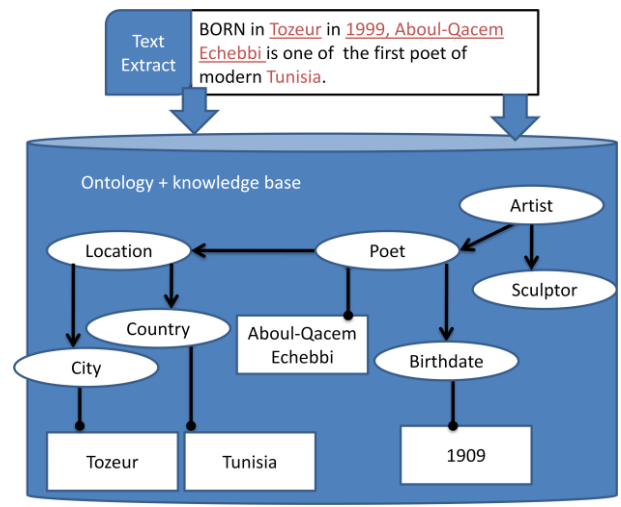

**Figure**. 3 The relation between resources in RDF graph and ontology.

In a nutshell, Semantic Annotation is about assigning to the entities in the text links to their semantic descriptions (as presented in Figure 3). Several types of annotation models are accessible in the literature and in the existing end-user applications. The models of semantic Web community have been abstracted by the Subject-Object-Predicate triple usefully for most of the annotation kind discussed here.

## 3.1 Manual Annotation

Manual annotation (MA) is a methodology that transforms the existing syntactic resources into interlinked knowledge structures by adding information to some level of document (word, phrase or paragraph) which constitutes metadata. The process of manual annotation is expensive, and regularly does not consider that multiple standpoints of a data source, involving multiple ontologies, can be useful to support the requirements of different users. Furthermore, MA is more easily feasible today, by means of authoring tools such as Semantic Word. MA is more precise compared to automatic annotation, but is very labor-intensive. As an example of manual annotation, in Protégé, a user can create an instance each time he wants to annotate by the selection of a piece of text in the loaded document and the class selection from the loaded ontology. As another example, in clinical documents, the instances that occur at different locations in the document actually refer to the same real world entity. Such annotations on equivalences can be valuable to deduce new knowledge and useful for medical care related applications [7]. The instance relationship annotation is another important type of annotation. Semantator [8] is an example of a tool that allows users to create a single relationship between two instances at a time. Semantator allows a given user to select two instances and add them to the relationship candidate list. After that, he can choose any object property from the loaded ontology and decide the subject of this new relationship. Another important type of annotation is instance relationship annotation allowing users to create a single relationship between two instances at a time.

## 3.2 Semi-automatic Semantic Annotation

The semi-automatic annotation process needs human intervention at some annotation level. This category of annotation systems differs in their architecture, methods and tools of information extraction, the manual work amount required to achieve annotation, performance, storage management and other features. GATE is an example of tools that performs semi-automatic annotation. The NCBO annotator [9] and cTAKES [10] are two other tools using semi-automatic annotation, in addition to Semantator. Similarly to that supported by NCBO annotator, cTAKES tool is used in Semantator for semi-automatic annotation. In a different manner to NCBO annotator, cTAKES is designed for clinical domain, uses NLP techniques and supports negation and time constraints. Currently, cTAKES allows annotation with the SNOMED CT and RxNorm dictionaries.

### 3.3 Automatic Semantic Annotation

Automated annotation of web documents is an important task in the Semantic Web effort. Semantic metadata created using automated annotation or tagging tools accompanied with best results are built on various machine learning algorithms which need training sets. Automated annotation tools can afford semantic metadata for semantic web and knowledge management [11]. Automatic semantic annotation can be realized on the base of automatic annotating algorithms: such as PANKOW (Pattern-based Annotation through Knowledge On the Web) and C-PANKOW (Context-driven and Pattern based Annotation through Knowledge on the Web) [12] for texts using Qtag[5] patterns and Google API. Additionally, Automatic semantic annotation can be based on statistical algorithms for image and video annotations. Nevertheless, current annotations based on automatic algorithms need to be improved and corrected. As an interesting tool for automatic semantic annotation, AeroDAML [13] loads a specific ontology and automatically produces the corresponding semantically marked up page which can then be checked by a human. KIM[6] [14], for example, is an automatic semantic annotation platform which uses information extraction based on GATE[7] [15] information extraction system, GATE with Annie[8] extension. SemTag [16] is the distinguished semantic annotation solution that operates within the distributed architecture and capability to process large scale data. To support annotation tasks, SemTag uses the Seeker [16] information retrieval platform and annotate web pages using Stanford TAP ontology [17]. Additionally, SemTag allows identifying but not creating new instances in the ontology. Ontea [18] is a platform for pattern based automated semantic annotation which allows pattern implementation over regular expressions, the implementation or integration of other methods based on patterns such as wrappers, solutions using structure of document, XPath, language patterns, e.g. C-PANKOW or GATE.

## 4. SEMANTIC ANNOTATION DOMAINS AND MODELS

### 4.1 Semantic Annotation Models

Semantic Annotation may be classified into four models: tags, attributes, relations and ontologies. Tags are located at the bottom level and correspond to the easiest form of annotation from the user point of view; while ontologies are at the top level and represent the hardest form of annotation from the user point of view.

- **Tags:** A tag annotation element is a keyword (word or sequence of characters without spaces) or a term assigned to a resource that, implicitly, describes a particular property of a resource. The annotation meaning planned by the annotator is not known by the computer and other users, only if the natural language used is unambiguous. The place names where picture are taken, the name of the person on a picture, or the topic of a news article are examples of tag.
- **Attributes:** An attribute annotation element is a pair of two elements: the name of the attribute and the value of the attribute. The name of the attribute defines the property of the annotated resource (e.g.,

---

[5] http://www.english.bham.ac.uk/staff/omason/software/qtag.html
[6] http://www.ontotext.com/kim/semanticannotation.html
[7] http://gate.ac.uk/
[8] http://gate.ac.uk/ie/annie.html

"Country", "birthdate") and the attribute value specifies the corresponding value (e.g., "Tunisia", "1909").

- **Relations:** a relation annotation element is a pair of two components: the relation name and the related resource. The annotated resource is related with relation by the relation name. In other words, the model of relation annotation is an extension of the model of attribute annotation to the domain of resources, allowing the user to interconnect these resources. For example, a citation referencing another paper in given scientific paper is an example of the annotation of relation defining a link between these documents.
- **Ontologies:** The ontology model describes the metadata that align a resource or a part of it with some of its properties and characteristics description according to a formal conceptual model (ontology). As defined by Studer et al., "an ontology is an explicit specification of a (shared) conceptualization" [5]. Ontologies are useful for domain knowledge capturing (in a generic manner) and the specification of a commonly granted understanding of a domain (that may be reused and shared within communities or applications). The design of ontologies may be realized with the following elements: concepts notion, concepts instances, concepts and instances properties, restrictions on these properties, relations between concepts and relations between instances. The user that use ontology annotation model is able to describe and connect existing resources by the resources structuring (concepts or as instances) and by the definition of the restrictions that hold between relations and properties.

## 4.2 Annotation Domains

Annotation domains were classified to document annotations, semantic wikis, semantic blogs, and tagging. A short introduction to each annotation domain with the specification of the associated role is presented in the following sections:

- **Document annotations:** Annotations of documents is the attachment of comments, notes, explanations, or other types of external remarks to a Web document or to a selected part of a document. Annotation of document can be realized manually (performed by user(s)), semi-automatic by automatic suggestions, or completely automatic.
- **Semantic Wikis:** Wiki can be defines as an environment for collaborative hypertext authoring that allows people collecting, describing, and authoring information in a collaborative manner. As a promising tools, semantic Wikis allow users to make formal descriptions of resources (wiki pages) enabling metadata insertions through semantic annotations and link relations between those resources. They need ontologies as conceptual models for their content organization. Annotations are needed to refer to an ontological model which defines concepts and properties associated to pieces of wiki contents.
- **Semantic Blogs:** A blog is a web site or online journal including comments, reflections and a lot of hyperlinks provided by the writer, but presented in reverse chronological order. However, the success of blogging can be reinforced when it accompanied with machine-readable content (annotation) which is beneficial of blog content consumers. Most commonly, an annotation in blogs is a statement about a post. For example, while a blog solution allows the classification of posts with simple categories or topics such as "research", "teaching"; we can say that blog posts are annotated with these categories. The process that transform blogs from simple online record to full participants in an information sharing network exploiting the metadata richness is called semantic blogging.
- **Tagging:** Organizing electronic content in a collaborative form by marking content with descriptive terms (tagging by keywords or tags) is a common way for future navigation, filtering or search. The tags express some undetermined relation between the resource and whatever the term refers to. As an example, del.icio.us[9], Technorati[10] or Flickr[11] are three tagging systems allowing users to associate one or more tags to a web resource.

## 5. ANNOTATION TOOLS

Annotation metadata can have several formats (textual, ontological, image or multimedia). The following paragraphs give a detailed description for each tool format.

## 5.1 Ontology-based annotation tools

Based on the criteria of tools that capture the requirement of providing explicit formal meaning to annotations, the following tool was selected:

- **Kim Plugin** [19] (Sirma Inc.): **KIM** platform is a part of the SWAN (Semantic Web ANnotator) project. It is a fully automatic and unsupervised tool for semantic annotation which works with its own meta-ontology.
- **Melita** [20] (University of Sheffield): Melita is a tool developed to define and develop an ontology-based annotation services. It is a semi-automatic annotation tool based on the Amilcare Information Extraction Engine [21].
- **Ont-O-Mat** [22] (AIFB): Is an implementation of the S-CREAM, a framework that supports both manual and interactive semi-automatic annotation of texts. Ont-O-Mat uses an automated data extraction technique from Amilcare (an adaptive Information Extraction) system designed to support active annotation of documents.
- **MnM** [23] (KMI): In a similar manner to Melita and Ont-O-Mat, MnM tool provides both automated and semi-automated support, based on the Amilcare system that support ontologies formalized in RDF.
- **C-Pankow** [24] (AIFB)[12]: C-Pankow is a fully automatic and unsupervised tool for semantic

---
[9] https://delicious.com/
[10] http://technorati.com/
[11] http://www.flickr.com/
[12] Smore 5.0 was not evaluated due to lack of support of annotation metadata.

annotation that support ontologies formalized in RDF.

## 5.2 Image Annotation Tools

Semantic annotation of images necessitates multimedia ontologies. Several vocabularies can be exploited (Dublin Core, FOAF), but they don't provide suitable models to describe sufficient multimedia content for sophisticated applications. The following section gives some example of image annotation tools:

- **ALIPR** [25] is a real-time automatic image tagging engine system fully automatic and high speed annotation for online pictures. It annotates images based on content.
- **GIAM** (Generalized image annotation methods) [26] [27][28] are designed to be used across a large number of images but needs a high intra-category clustering with adequate intercategory separation. Because the search space grows, categorical separation becomes challenging with GIAM.
- **SIAM** (Specialized image annotation methods): Conversely to GIAM, specialized annotators frequently perform well within their domains [29][30], but need a-priori assumptions about the data which, for a general image set are incorrect.
- **SpiritTagger** [31] : Is a system of image annotation invented to explore knowledge extraction through mining of millions of global photographs referenced with a geographical coordinate.
- **SML** (supervised multiclass labeling) [32]**:** it also 1) produces a natural ordering of semantic labels at annotation time, and 2) eliminates the need to compute a "nonclass" model for each of the semantic concepts of interest.
- **CRM (**Continuous-space Relevance Model**) [33]:** CRM is an image annotation and retrieval tool based on probabilistic model. It is designed to reduce an image to a real-valued feature vectors set, and subsequently model the joint probability of observing feature vectors by means of potential annotation words.
- **MBRM** (Multiple Bernoulli Relevance Models)[33]: MBRM is based on the CRM model presented above and uses a multiple-Bernoulli distribution for modeling image annotations over CRM.

## 5.3 Text Annotation Tools

There is a good number of textual information that would be more useful if it were annotated for the Semantic Web, but the nature of the data makes it difficult to do so. As examples of this type of data are the EBay posts texts, internet classifieds like Craig's list, bulletin boards such as Bidding for Travel, or even the text summary below the hyperlinks returned after querying Google. As example of tools:

- **Amaya** [34]: Amaya is an annotation tool allowing user to make annotations in the same tool they use for browsing and for editing text by mark-up Web documents in XML or HTML. It is a good example of a single point of access environment. Amaya manual annotation of web pages is allowed, but requires the features to support automatic annotation. All Amaya annotations may be realized by **Annozilla**[13] browser aiming to make readable in the shadow Amaya developments and the Mozilla browser.
- **AktiveDoc** [35]: AktiveDoc is a client-server application integrated in a Web Based KM system. It allows annotation of documents at three levels: free text statements, on-demand document enrichment and ontology based content annotation. AktiveDoc provides Semi-automatic annotation of content based on Amilcare. AktiveDoc is able to provide automatic suggestions about relevant content, given its design for knowledge reuse. ActiveDoc functionality is extended to free text, in addition to filling forms functionality and other pre-determined structures.
- **Magpie** [36]: Magpie is a real-time annotation of web resources that relates text strings to ontology concepts of the user's choice. With Magpie, an appropriate web service can be linked to highlighted strings. Even though, the annotation of documents is automatic, Magpie has the disadvantage to produce manually the lexicons specific parts of text strings subjects for each ontology. In Magpie, the work on automating lexicon generation is in progress.
- **Thresher** [37]**:** Is a similar system to Magpie (similarity in the use of wrappers that generates RDF on the fly as users browse deep web resources) that lets non-technical users teaching their browsers semantic web content extraction from HTML. Additionally, Thresher allows the user to access semantic services for recognizing objects. Since Thresher is part of the Haystack semantic browser [38], users can also do the personalization of the ontology that use.
- **WiCKOffice** [39]: WiCKOffice provides annotation for word processor files, similar to OntoOffice[14]( A commercial annotation system for Microsoft Office applications available from Ontoprise). It helps in filling forms using data extracted from knowledge bases and demonstrates the effectiveness of writing within a knowledge aware environment to support possibilities (automatic assistance for form).

## 6. CONCLUSION

This paper has presented a number of issues related to the representation and the usage of the semantic annotation. Firstly, it describes some concepts of the Semantic Web, including metadata annotation which helps to make Semantic Web vision a reality. Additionally, this paper has presented three classes semantic annotation with some recognized systems (manual, semi-automatic and automatic). The domains, tools and the models of semantic annotations are also described. In future we plan to evaluate the described approaches of semantic annotation based in several criteria and different domains.

---

[13] Annozilla annotator (http://annozilla.mozdev.org/index.html accessed on 3 August 2004).

[14] OntoOffice tutorial (http://www.ontoprise.de/documents/tutorial ontooffice. pdf accessed on 30 November 2004).